\documentclass[twocolumn,noshowpacs,noshowkeys,pra,aps,superscriptaddress,bibliography]{revtex4-1}%
\usepackage{color,graphicx}
\usepackage{pgfplots}
\usepackage{braket}
\usepackage{amsmath}

\begin{document}

\title{Optimised photonic crystal waveguide for chiral light-matter interactions}

\author{Ben Lang}
\email{bl9453@bristol.ac.uk}
\affiliation{Quantum Engineering Technology Labs, H. H. Wills Physics Laboratory and Department of Electrical \& Electronic Engineering, University of Bristol, BS8 1FD, UK}

\author{Ruth Oulton}
\affiliation{Quantum Engineering Technology Labs, H. H. Wills Physics Laboratory and Department of Electrical \& Electronic Engineering, University of Bristol, BS8 1FD, UK}

\author{Daryl M. Beggs}
\affiliation{School of Physics and Astronomy, Cardiff University, Queen's Buildings, The Parade, Cardiff CF24 3AA, UK}

\begin{abstract}
We present slow-light photonic crystal waveguide designs that provide a $\times$8.6 improvement of the local density of optical states at a fully chiral point over previous designs.
\end{abstract}

\maketitle

\section{Introduction}

Chirality of light in (nano-)photonic structures is proving to be a valuable resource \cite{chiral_quantum_optics,nano_anntenna}.  In quantum optics, chirality couples the spin direction of electrons to the travel direction of light.  This chiral light-matter interaction is at its most useful when the chirality reaches 100\% at a singular position known as a C-point.  A quantum dot (QD) placed at a C-point can display a spin-dependent unidirectional-emission \cite{young_prl} – an attractive property for quantum optics, as it allows spin-encoded static qubits to be converted to path-encoded flying qubits. 

Photonic crystal waveguides (PhCWGs) present several unique benefits to realising chirality-direction coupling. Firstly, the longitudinal component of the waveguide modes is large, meaning that C-points with 100\% chirality are common, and moreover tend to occur in the high-index part of the waveguide, where a quantum dot could be placed. Secondly PhCWGs support slow-light.  Extending the benefits of slow-light to the interaction of a quantum dot at a C-point is attractive.  Slow-light enhances the density of optical states allowing extremely bright sources and high collection efficiencies to be realised \cite{efficient}.  In principle, there is no upper limit to the light-matter interaction enhancement.  At the bandedge, the PhCWG has stopped modes with group velocity $v_g = 0$ and an infinite density of optical states (the van Hove singularity \cite{Van}). Only the practicality of fabricating a perfect waveguide without defects prevents the use of these bandedge modes.

As shown by our recent work, the case of an emitter placed at a C-point requires additional consideration.  Time-reversal symmetry dictates that all chiral components of the waveguide mode must vanish in the stopped light at the bandedge, and therefore no C-points can be supported.  Modes in the slow-light regime resemble the bandedge mode, and thus contain less chirality. This forces a compromise between strong light matter interactions (for which slow light is desirable, typically found near the bandedge) and making those interactions chiral (for which we want powerful circularly polarised fields, which become scarce near the bandedge).  We found that the optimum local density of optical states (LDOS) at a C-point is found in modes with modest group velocities of $v_g < c/10$ for the standard PhCWG design (the so-called W1 waveguide) \cite{bandedge_us}. Only a limited number of alternative nano-photonic designs for quantum dots have so-far been considered \cite{glide1, Coles_direction}. In this paper we present alternative PhCWG designs that possess larger LDOS at the C-points.


Our search for these designs begins with the archetypical W1 waveguide of one row of missing holes from an hexagonal lattice of holes with radius $r = 0.3a$ in a GaAs dielectric membrane.  We form new designs by displacing each hole in the first row of holes closest to the waveguide a distance $D1$ towards the waveguide core, and the second row of holes a distance $D2$ (fig.1a).  In this way the dispersion \cite{disp_eng} and electric field profile \cite{field_eng} of the PhCWG can be modified.  Modifying the hole positions is preferred to the radii as they are more accurately realized in electron-beam lithography \cite{design_review}.

\begin{figure}
\includegraphics[scale=0.33]{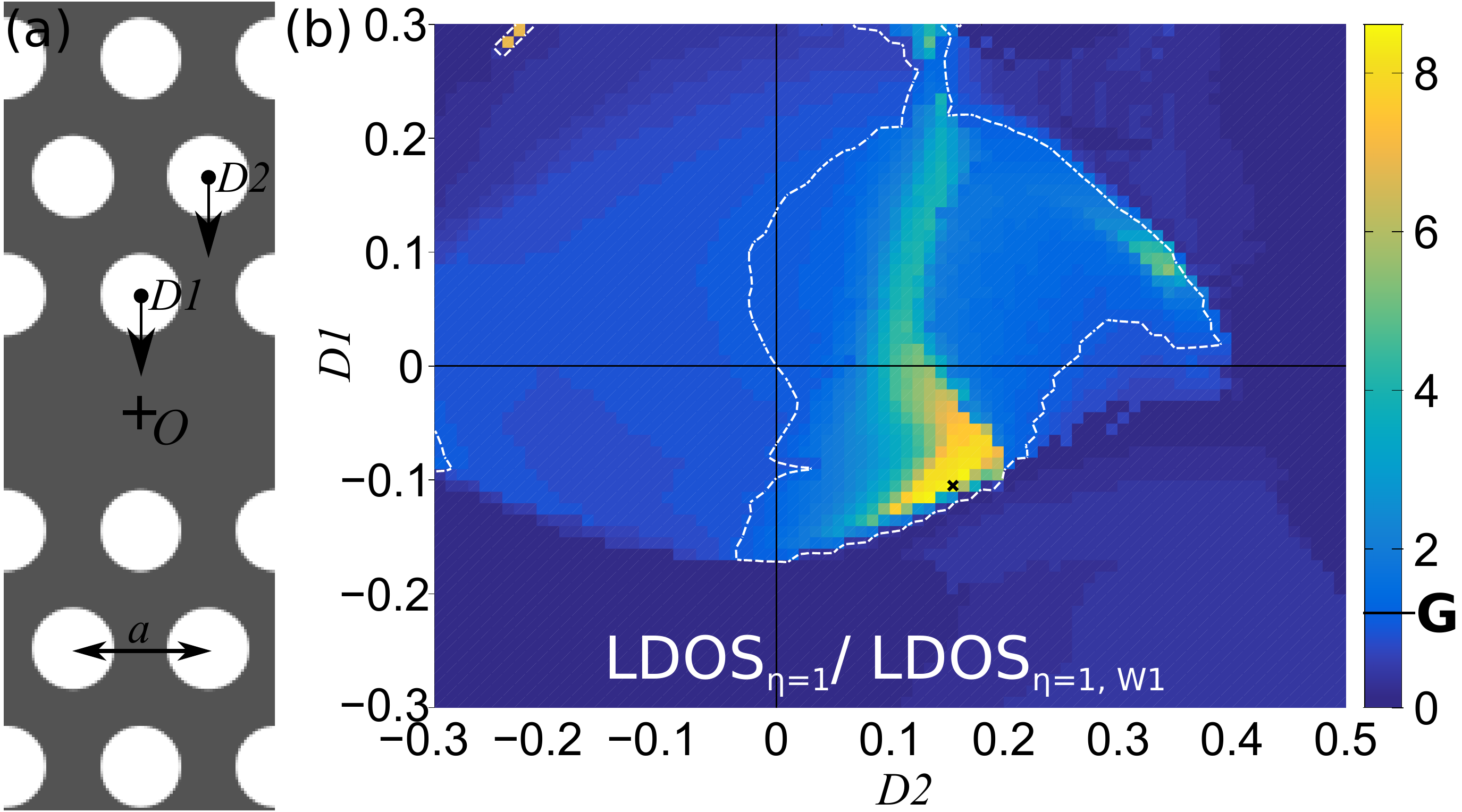}
\caption{Fig. 1. (a) The refractive index profile of a W1. The arrows indicate how $D1$ and $D2$ control the displacements of the innermost holes. $O$ marks the origin of the in-plane coordinate system $(x,y)=(0,0)$. (b) Largest LDOS at a C-point as a function of $(D1,D2)$, normalized to the largest LDOS at a C-point in a W1.  The dotted line is a contour along which this ratio is unity. {\bf G} on the colour bar indicates the value for the glide waveguide described in ref. \cite{glide1}.}
\label{fig1}
\end{figure}

Figure 1b presents the main results, showing the enhancement of the LDOS at C-points where they occur in a single-mode waveguide.  The vast majority of designs show little or no enhancement over the standard W1 (i.e. the ratio of LDOS at the C-points is below or close to one), but there is a small region of the search space that show significant enhancements.  Of these, the best design identified has 8.61 times  the C-point LDOS of a W1 for $(D1, D2) = (-0.11, 0.15)a$ at a C-point at frequency $0.2791c/a$ and position $(x, y) = (0.5, 1.170)a$ from the origin shown in fig.1a.

Finite difference time domain (FDTD) simulations confirming the unidirectional emission from the best identified C-points in the optimized and W1 waveguides are shown in fig.2a \cite{meep}.  Figure 2b shows the calculated power radiated in the forwards, backwards and sideways directions.  The $\sim \times10$ enhancement in LDOS is well replicated in these calculations.  Emission into the backwards direction is suppressed by a factor of $10^6$ in the W1 and $10^4$ in the optimized design.  The lower directionality in the optimized waveguide is due to differences in the bandwidth over which the PhCWGs show strong directional behaviour at these dipole locations.

The LDOS is a function of position and frequency and is proportional to the product $n_g |E|^2$, where $n_g = |c/v_g|$ is the group index and $|E|^2$ is the electric field intensity.  We are interested in positions of unit directionality, where the directionality is defined as the difference between power emitted by a spin transition (modelled as a circular point-dipole) into the forwards and backwards waveguide modes, normalized by the total power emitted into the waveguide.  For a single-mode waveguide the directionality is simply given by the degree of chirality  $\eta = |S_3|= 2 |Im(E_x^* E_y )|/|E|^2$.  $\eta=1$ is typically not possible if the waveguide is multi-mode (it requires the extraordinary coincidence of C-points at the same position in all modes at the same frequency) and so multi-mode frequency regions are ignored.

For each choice of $(D1,D2)$ we search for C-points with $\eta=1$, and calculate the relative LDOS ($n_g |E|^2$) at each using a frequency domain eigensolver \cite{mpb}.  In each design $(D1,D2)$, the C-point with the highest LDOS is compared to the C-point with the highest LDOS in the W1 design $(D1,D2)=(0,0)$. Our calculations use the effective index method \cite{effective_index}, which allows 2d simulations to approximate 3d slab structures.  We treat values of $n_g >100$ as $100$ in the LDOS calculation, serving the dual purpose of focusing the search on experimentally achievable values \cite{ng_100} and filtering computational errors at the bandedge where $|E|^2 \rightarrow 0$ and $n_g \rightarrow \infty$.

\begin{figure}
\includegraphics[width=0.5 \textwidth]{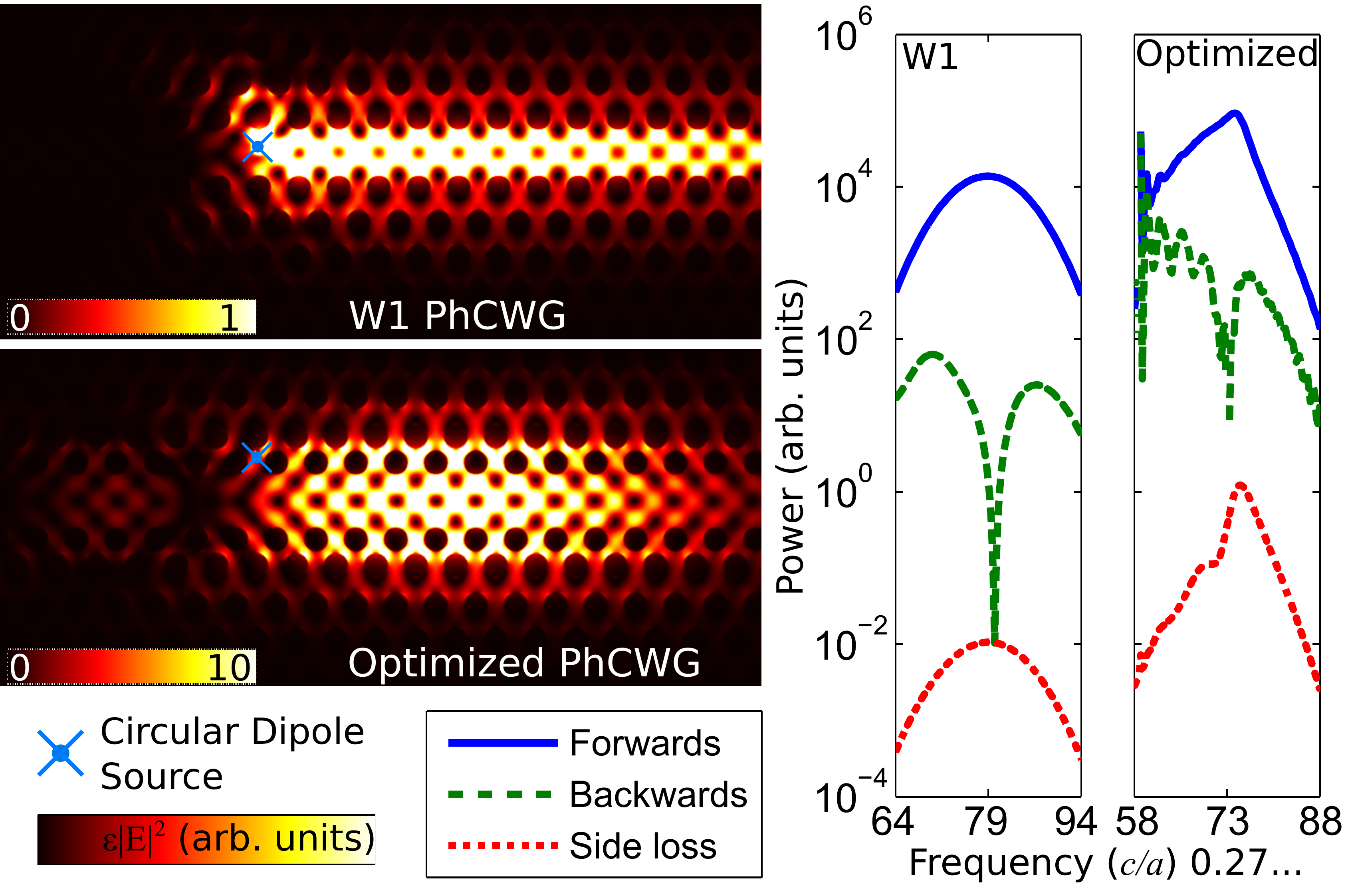}
\caption{Fig.2. (a) 2d FDTD simulations of circular dipoles at the ideal points in the optimized and W1 waveguides. (b) Power radiated forwards, backwards and sideways: note the log scale.}
\label{fig2}
\end{figure}

The best designs presented here exploit slow light far from the bandedge by flattening the dispersion. There is an element of “brinkmanship” to this, as flattening the dispersion towards an inflection-like curve maximizes the LDOS, but overshooting creates a multimode region that ruins the performance. This abrupt drop is visible in fig.1b on the lower side of the bright region.

This motivates a minor digression. In principle a perfect inflection point in the dispersion appears able to support uni-directionality with an infinite density of states. However even infinitesimal perturbations (for example, from disorder) can break the inflection point into a local maximum and local minimum pair with a small separation. This local maximum/minimum pair results in a multi-mode region with a slow-light mode allowing (strong) emission in both directions. Although such abruptness may at first glance seem un-physical, it is a consequence of us considering only modes with real $k_x$ (propagating modes), equivalent to considering an infinitely long PhCWG. Evanescent modes in a finite PhCWG always allow some light to tunnel from the dipole out of the waveguide in the wrong direction. Near an inflection point there exist weakly evanescent modes (with small imaginary $k_x$) that can tunnel a long distance \cite{inflection}. These weakly evanescent modes smooth out this transition in finite-length PhCWGs.

In real-world samples, QDs are typically strain-grown in random locations \cite{random_locations} (although positioning methods are being developed \cite{position_control}).  Furthermore, the size and shape dispersion means that the resonant frequencies of the QDs are randomly distributed around the desired one.  Experiments then typically proceed by testing a large number of samples, until a suitable one is found.

In the above, we have calculated the performance of the PhCWG for an ideally placed QD pitched at the ideal frequency, but we are also interested in the yield: how many samples can we expect to test before finding a good QD positioned at or near a C-point.  To answer this question we have also calculated the probability that a QD placed at a random location in the GaAs with a random frequency (selected uniformly from the bandwidth of the fundamental mode) will have $\eta > 0.9$. Our calculations neglect positions with negligible LDOS and assume $\eta < 0.9$ at multimode frequencies.

Fig.3 shows the result of this calculation. The optimized waveguide identified above (white cross, fig.3) has a poor yield ($\sim1$\%).  However by consideration of Fig. 1b and Fig.3 together PhCWGs with a desired compromise of yield and performance can be chosen. For example in the region $(D1,D2)\approx(0.1,0.15)$ both yield and performance are high.

\begin{figure}
\includegraphics[scale=0.33]{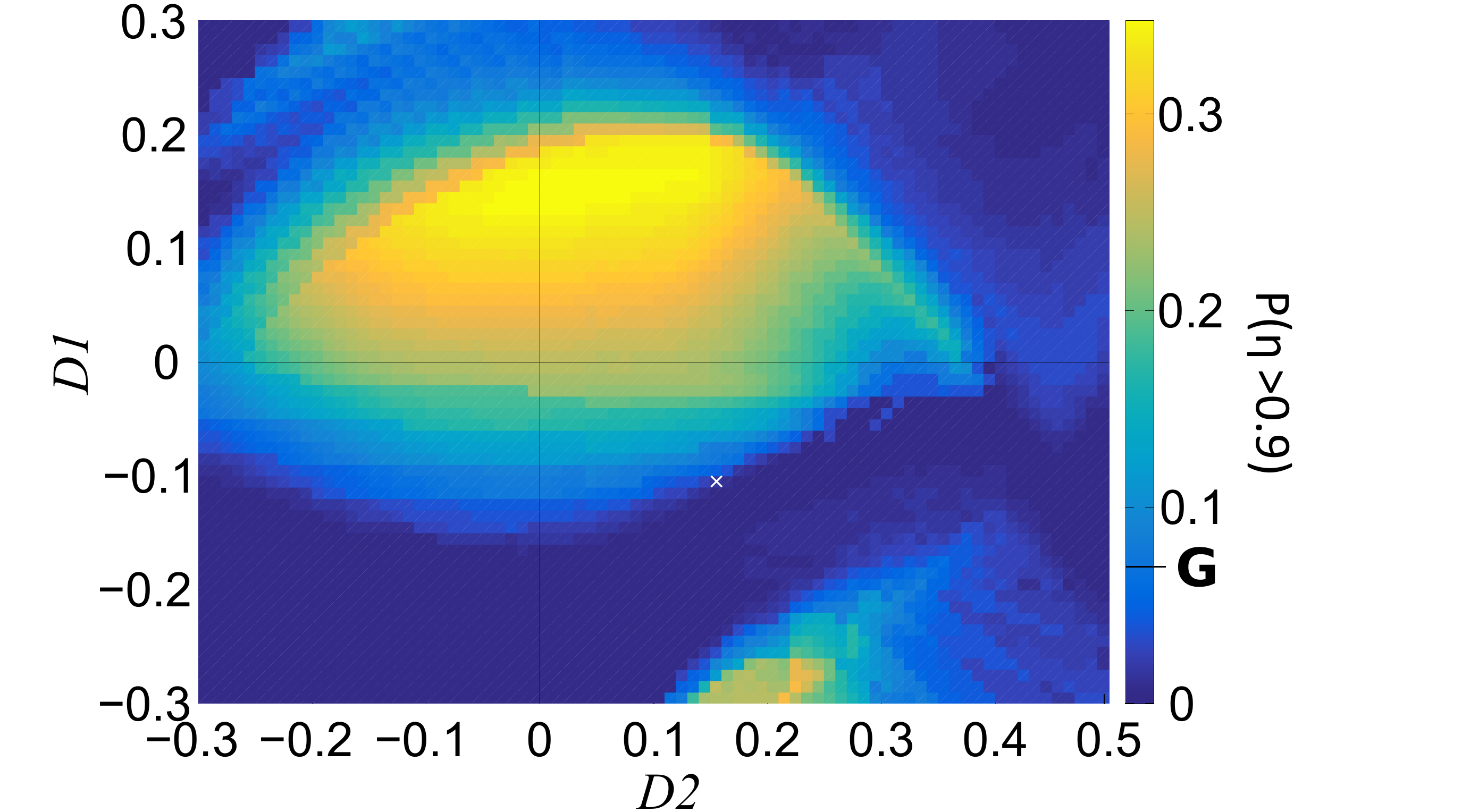}
\caption{Fig.3. Calculated yield.  The probability that directionality $\eta$ exceeds 0.9 for randomly placed QDs. {\bf G} has the same meaning as in fig.1.}
\label{fig3}
\end{figure}

In conclusion we have identified modified PhCWG designs that promise significant increases in chiral performance, with a $\times$8.6 enhancement of the LDOS.  As the LDOS is a measure of the light-matter interaction strength, and is directly proportional to the emission rate and efficiency [16], our optimized design will allow fabrication of waveguides almost one order of magnitude brighter than using the standard W1 design.

In the final stages of this work we became aware of related work suggesting a modified glide-plane waveguide design for strong chiral interactions \cite{glide2}. Our calculations suggest this design is excellent with $>100$ times LDOS enhancement at the best C-point compared to a W1 (\emph{cf} fig.1) and a $\sim 15$\% yield (\emph{cf} fig.3).

\begin{acknowledgments}

SPANGL4Q: FP7-284743. RO: EPSRC, EP/G004366/1. DMB: Marie Curie individual fellowship, QUIPS. GW4 

This work was carried out using the computational facilities of the Advanced Computing Research Centre, University of Bristol http://www.bris.ac.uk/acrc/.

\end{acknowledgments}

\end{document}